\documentclass[a4paper]{JAC2003}
\usepackage{graphicx}
\usepackage{amsmath,amssymb}
\usepackage{cite}

\usepackage{varwidth}
\usepackage{xcolor}

\begin{document}
\title{OPERATIONAL EXPERIENCE WITH CRAB CAVITIES AT KEKB}

\author{Y.~Funakoshi for the KEKB Commissioning Group\\
KEK, Ibaraki, Japan}

\maketitle

\begin{abstract}

KEKB was in operation from December 1988 to June 2010. The crab cavities were installed at KEKB in February 2007 and worked very stably until the end of KEKB operation. Operational experience of the crab cavities with beams is described.

\end{abstract}

\vspace{-3mm}

\section{KEKB B-FACTORY}

KEKB B-Factory \cite{DR} was an energy-asymmetric double-ring e$^+$e$^-$ collider at KEK  in operation from December 1998 to June 2010. KEKB was operated mainly  at the $\Upsilon(4\textrm{S})$ resonance. It was composed of the low-energy positron ring (LER) operated at
3.5~GeV, the high-energy electron ring (HER) operated at 8~GeV, and
an injector linac. Two beams collided at the physics detector called `Belle'.
The machine parameters of KEKB with the crab cavities are listed in Table~\ref{t1} together with the design parameters.
 The highest luminosity, $2.108\times 10^{34}$~cm$^{-2}\,$s$^{-1}$, was achieved in June
2009. The peak luminosity is twice as high as the design value and is the  highest value in the world so far.

 \begin{table*}%1
\caption{Machine parameters of KEKB (27 June 2009). Parameters in parentheses are the design parameters.}
\label{t1}
\begin{center}
\begin{tabular}{ l  c  c  c }
\hline\hline
  & LER & HER &  Unit \\  \hline
Energy & 3.5 & 8.0 & GeV \\
Circumference &  \multicolumn{2}{c}{---\,3016\,---} & m \\
RF frequency &  \multicolumn{2}{c}{---\,508.88\,---} & MHz \\
Horizontal emittance & 18 (18) & 24 (18) & nm \\
Beam current & 1637 (2600)  & 1188 (1100)  & mA \\
Number of bunches & \multicolumn{2}{c}{---\,$1585^{\rm a}\,({\sim}\,4600^{\rm b}$)\,---} & \\
Bunch current & 1.03 (0.57) & 0.75 (0.24) & mA \\
Bunch spacing & \multicolumn{2}{c}{---\,1.84 (0.59)\,---} & m \\
Total RF voltage & 8.0 (5--10) & 13.0 (10--20)  & MV \\
Synchrotron tune $\nu_s$ &$-0.0246$ ($-0.1$ to $-0.2$) & $-0.0209$ ($-0.1$ to $-0.2$) & \\
%
%Beatatron tune $\nu_x / \nu_y$ & 45.506/43.561 (45.52/45.08) & 44.511/41.585 (47.52/43.08) & \\ \hline
Horizontal tune  $\nu_x $ & 45.506(45.52) & 44.511 (47.52) & \\
Vertical tune  $\nu_y$ & 43.561 (45.08) & 41.585 (43.08) & \\
Betas at IP $\beta_x^* / \beta_y^*$ & 120/0.59 (33/1) & 120/0.59 (33/1) & cm \\
Momentum compaction $\alpha$ & 3.31 (1--2) &  3.43  (1--2) & $\times 10^{-4}$\\
%Beam-beam parameters & 0.127/0.129 (0.039/0.052) & 0.102/0.090 (0.039/0.052) & \\ \hline
Beam--beam parameter $\xi_x$ & 0.127 (0.039) & 0.102(0.039) & \\
Beam--beam parameter $\xi_y$ & 0.129 (0.052) & 0.090 (0.052) & \\
Vertical beam size at IP $\sigma_y^*$ & 0.94$^{\rm c}$ (1.34) &  0.94$^{\rm c}$ (1.34) & $\rm \mu m$ \\
Beam lifetime & 133@1637 & 200@1188 & min@mA \\
Luminosity (Belle CsI) & \multicolumn{2}{c}{---\,2.108 (1.0)\,---} & $10^{34}$~cm$^{-2}$ s$^{-1}$ \\ \hline
Total integrated luminosity & \multicolumn{2}{c}{---\,1041\,---} & $\rm fb^{-1}$ \\ \hline\hline
\multicolumn{4}{l}{\small $^{\rm a}$\,With 5\% bunch gap.} \\
\multicolumn{4}{l}{\small $^{\rm b}$\,With 10\% bunch gap.} \\
\multicolumn{4}{l}{\small $^{\rm c}$\,Value estimated  from the luminosity, assuming that the horizontal beam size is equal to the calculated value.}
\end{tabular}
\end{center}
\end{table*}

The HER beam current exceeded the design value, but the LER beam current was lower than the design. This is not attributable to hardware limits; the luminosity saturated at around 1.6~A, and a higher beam current did not bring a higher luminosity. We believe that this is due to electron cloud instability. The bunch spacing is also much longer than the design, to mitigate the electron cloud instability. As a result, the bunch currents were much higher than the design. The vertical beta function at the interaction point (IP), $\beta^*_y$, was 5.9~mm, much lower than the design value. Because of the crab cavities, the vertical beam--beam parameter ($\xi_y$) was as high as 0.09, much higher than the design. Another feature of KEKB is that the horizontal
tune was very close to a half-integer; this also contributed to the high luminosity.  The daily integrated luminosity was twice as high as the design because of the continuous injection
mode and the acceleration of two bunches per radio-frequency (RF) pulse at the linac.

\begin{figure*}[th]%1
\centering
\includegraphics*[width=158mm]{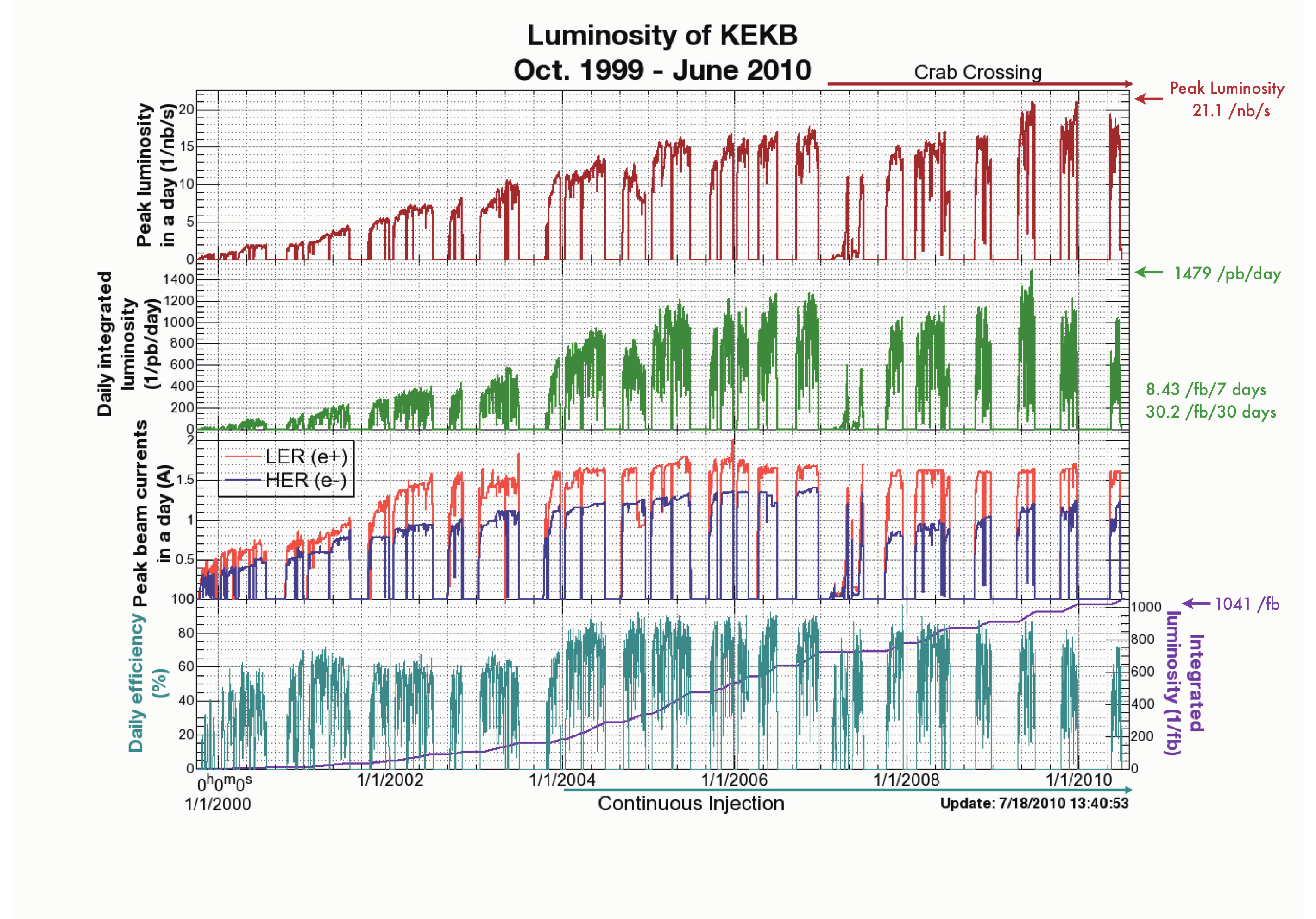}
\vspace{-10mm}
\caption{History of KEKB.}
\label{f1}
\vspace{0mm}
\end{figure*}

Figure~\ref{f1} shows the history of KEKB. The crab cavities were installed at KEKB in February 2007 and  worked stably until the end of KEKB operation. After installation of the crab cavities, the luminosity was somewhat lower than before the crab cavities were installed. Although the specific luminosity was higher, the beam currents, particularly in HER, were much lower and the luminosity was also lower. This was not due to a hardware limitation; as described below, it was caused by the dynamic beam--beam effects. Upon overcoming this problem, the luminosity increased. In addition, the skew-sextupole magnets, which were installed in the winter shutdown of 2009, contributed to a higher luminosity.

\begin{table}[ttt]%2
\tabcolsep 3pt
\begin{small}
\begin{center}
\caption{Comparison of KEKB machine parameters before and after installation of crab cavities.}
\vspace{2mm}
\begin{tabular}{l cc cc c}
\hline\hline
& \multicolumn{2}{c}{{May 2008}} & \multicolumn{2}{c}{{Nov 2006}} & \\
               &  LER       &   HER         &  LER      &     HER         &             \\ \hline
Energy         & 3.5        &   8.0         &  3.5      &     8.0         &   GeV       \\
Circum.  & \multicolumn{2}{c}{---\,3016\,---} & \multicolumn{2}{c}{---\,3016\,---}  &   m         \\
$\rm \phi_{cross}$& \multicolumn{2}{c}{Crab crossing} & \multicolumn{2}{c}{---\,$\rm 22$\,---}  &   mrad         \\
$I_{\rm beam}$ & 1619       &   854        &  1662     &  1340           &   mA        \\
$N_{\rm bunches}$  & \multicolumn{2}{c}{---\,1584\,---} & \multicolumn{2}{c}{---\,1387\,---}  &             \\
$ I_{\rm bunch}$& 1.02       &   0.539        &  1.20     &  0.965          &   mA        \\
$\varepsilon_x$        &  15        &     24         &   18        &  24              &   nm        \\
$\beta_x^*$    &  90        &     90         &   59        &  56              &   cm        \\
$\beta_y^*$    &  5.9       &     5.9        &   6.5       &  5.9             &   mm        \\
$\sigma_y^*$  &  1.1       &     1,1        &   1.9       &  1.9             & $\mu$m     \\
$V_{\rm C}$     &  8.0       &     13.0      &   8.0       &  15.0           &   MV        \\
$\nu_x$           &  0.505     &     0.509      &   0.505    &  0.509           &             \\
$\nu_y$          &  0.567      &     0.596     &   0.534    &  0.565           &             \\
$\nu_s$          &  $-0.0240\;$      &     $-0.0204\;$     &   $-0.0246\;$    &  $-0.0226\;$           &             \\
$\xi_x$            &  0.099      &     0.119      &   0.117    &  0.070           &             \\
$\xi_y$            &  0.097      &     0.092      &  0.105    &  0.056           &             \\
Lifetime           &  94       &     158       &   110     &  180            & min.        \\
Luminosity    & \multicolumn{2}{c}{---\,16.10\,---}& \multicolumn{2}{c}{---\,17.12\,---} &  /nb/s      \\
Lum/day        & \multicolumn{2}{c}{---\,1.092\,---}& \multicolumn{2}{c}{---\,1.232\,---} &  /fb        \\
\hline\hline
\end{tabular}
\label{l2ea4-t1}
\end{center}
\end{small}
\end{table}

\section{CRAB CROSSING scheme}

%------------------------
\subsection{Motivation of Crab Cavities}
%------------------------

\begin{figure}[thh]%2
\centering
\includegraphics*[width=87mm,clip]{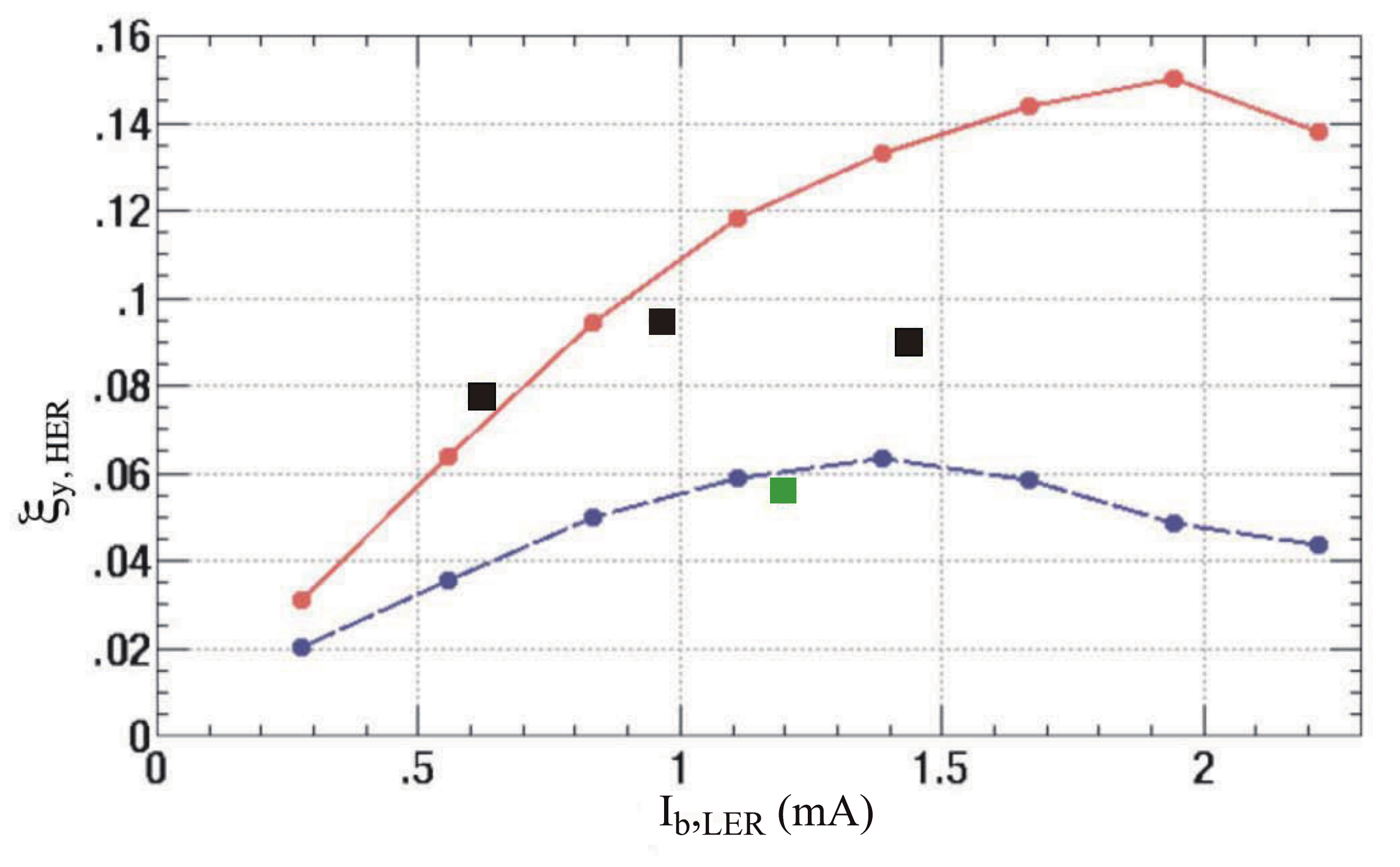}
\caption{Predicted beam--beam parameters obtained from strong--strong beam--beam simulations with  crossing angle  22~mrad (purple) and for head-on or crab crossing (red). Some experimental data are also shown (squares), with black and green squares representing data with and without the crab cavities, respectively.}
\label{fcrab1}
\end{figure}

One of the design features of KEKB is the horizontal
crossing angle of $\rm \pm 11$  mrad at the IP. Although the crossing
angle scheme has many merits, the beam--beam performance may degrade.
In the design of KEKB it was predicted that the vertical beam--beam
parameter  $\xi_y$ could be as high as 0.05 if betatron tunes are
 chosen properly.
The crab crossing scheme was proposed  by R.\ Palmer in 1988 \cite{RP}  as an approach to recovering the head-on collision with the crossing angle for linear colliders. It has  also been shown that the synchro-betatron coupling terms associated with the crossing angle in ring colliders are cancelled by crab crossing \cite{OY}. The crab crossing
scheme has been considered in the design of KEKB as a back-up measure to guard against possible problems with the crossing angle.
Previously the crab cavities had seemed not to be urgently necessary, as KEKB achieved  $\xi_y > 0.05$ at the early stage of its operation in 2003. Later, however, interesting
beam--beam simulation results appeared \cite{KO,KO2,KO3}, predicting that
the head-on collision or crab crossing provides a higher  value of  $\xi_y$, around 0.15, if combined with a horizontal tune that is very close to a half-integer, such as 0.508.
Figure~\ref{fcrab1} shows the comparison of $\xi_y $ for the head-on collision (crab crossing)
with that for the crossing angle, obtained by a strong--strong beam--beam simulation.
Afterwards,  the development of crab cavities was revitalized, and they were finally installed at KEKB in February 2007.

%------------------------
\subsection{Single Crab Cavity Scheme}
%------------------------

\begin{figure}[hhh]%3
\vspace{3mm}
\centering
\includegraphics*[width=87mm,clip]{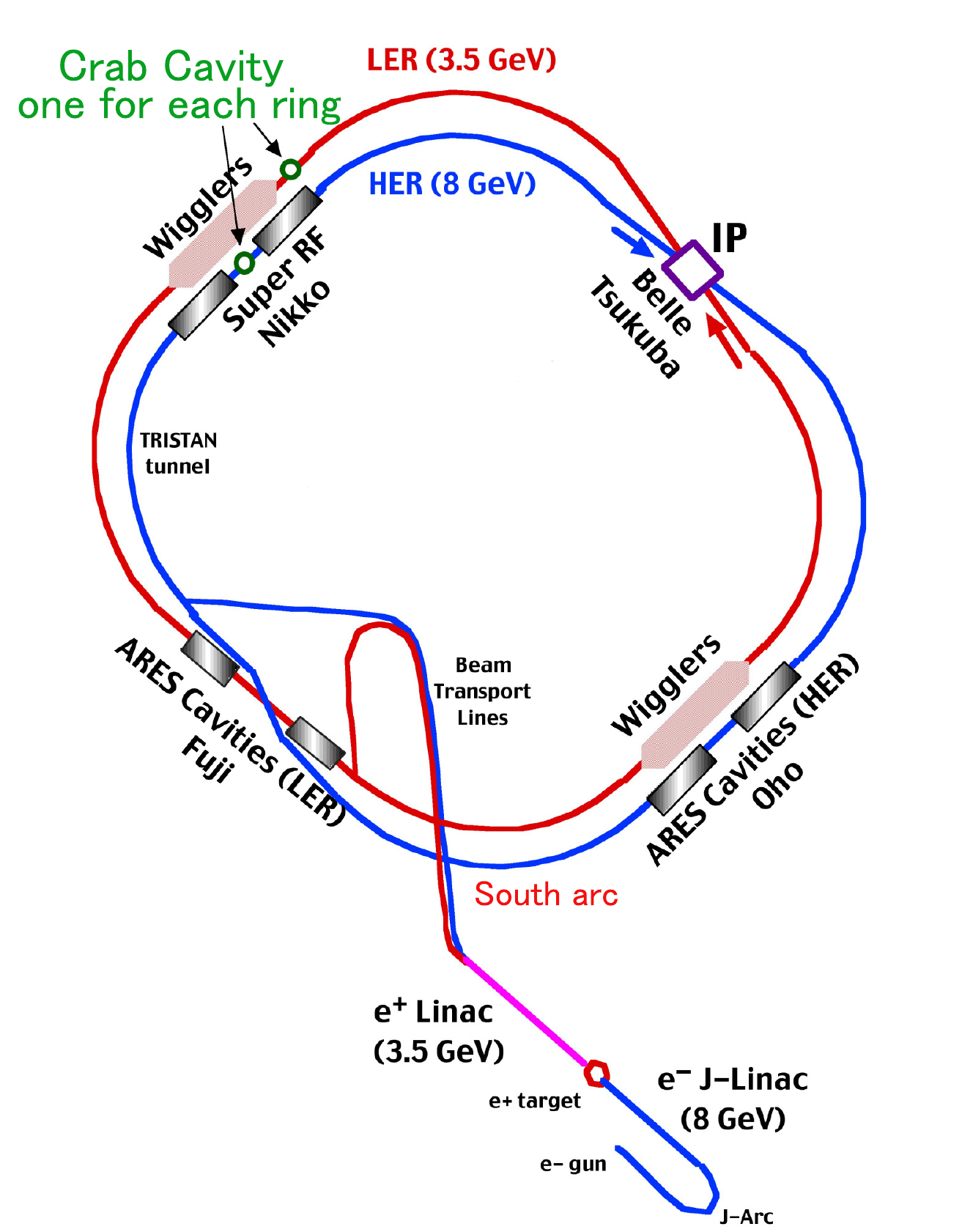}
\caption{Layout of  the KEKB rings and crab cavities.}
\label{fcrab2}
\end{figure}

In the original design of KEKB, we had planned to install two crab cavities for each
ring on both sides of the IP, so that the crab kick excited by
the first cavity would be absorbed by another one. The single
crab cavity scheme extends the region with crab orbit until
the two cavities eventually merge with each other at a particular
location in the ring. Thus it needs only one cavity per ring.
The layout is shown in Fig.~\ref{fcrab2}. This
scheme not only saved us the cost of the cavities but also made it
 possible to use the existing cryogenic system in the Nikko region,
which has been utilized for the superconducting accelerating cavities.

In the single crab cavity scheme, the following equation should be satisfied for the two beams to achieve a head-on collision:
\[
\frac{\phi_x}{2} = \frac{\sqrt{\beta_x^{\rm C}\beta_x^*}\,\cos\bigl(\pi\nu_x-|\Delta\psi_x^{\rm C}]\bigr)}{2 \sin\pi\nu_x}\frac{V_{\rm C}\,\omega_{\rm RF}}{Ec}.
\]
Here $\phi_x$ is the full crossing angle; $\beta_x^{\rm C}$ and $\beta_x^*$ are the beta functions at the crab cavity and the IP, respectively; $\Delta\psi_x^{\rm C}$ denotes the horizontal betatron phase advance between the crab cavity and the IP; $\nu_x$ is the horizontal tune; and $V_{\rm C}$ and $\omega_{\rm RF}$ are the crab voltage and the angular RF frequency, respectively. Typical values for these parameters are shown in  Table~\ref{table2}.

\begin{table}[hbt]%3
\begin{center}
\caption{Typical parameters for the crab cavities. The crossing angle, the horizontal beta functions at the IP and crab cavities, the horizontal tunes, the horizontal phase advance from the cavities to the IP, the crab voltage, and the RF frequency are shown.}
\vspace{2mm}
\begin{tabular}{lccc}
\hline
\hline
                             &  LER                                      &   HER                              &       Unit              \\ \hline

$\phi_{x}$                  &      \multicolumn{2}{c}{---\,$\rm 22$\,---}                   &    mrad       \\
$\beta_{x}^*$            &  1.2      &  1.2         &    m            \\
$\beta_{x}^{\rm C}$           &             51                     &            122                          &    m             \\
$\nu_{x}$                    &             45.506              &            44.511                  &                      \\
$\psi_{x}^{\rm C}/2 \pi$     &             0.25                   &            0.25                        &                     \\
$V_{\rm C}$                        &             0.97                   &            1.45                        &    MV            \\
$f_{\rm RF}$                        &             \multicolumn{2}{c}{---\,508.89\,---}                     &    MHz         \\
\hline
\hline
\end{tabular}
\label{table2}
\end{center}
\end{table}

The beam optics was modified for the crab cavities to
give the necessary magnitude of the  beta functions at the cavities
and the proper phase advance between the cavities and IP.
A number of quadrupoles have switched polarity and come to have independent power supplies.

%------------------------
\section{Operation with crab cavities}

\begin{figure}[hhh]%4
\vspace{3mm}
\centering
\includegraphics*[width=87mm,clip]{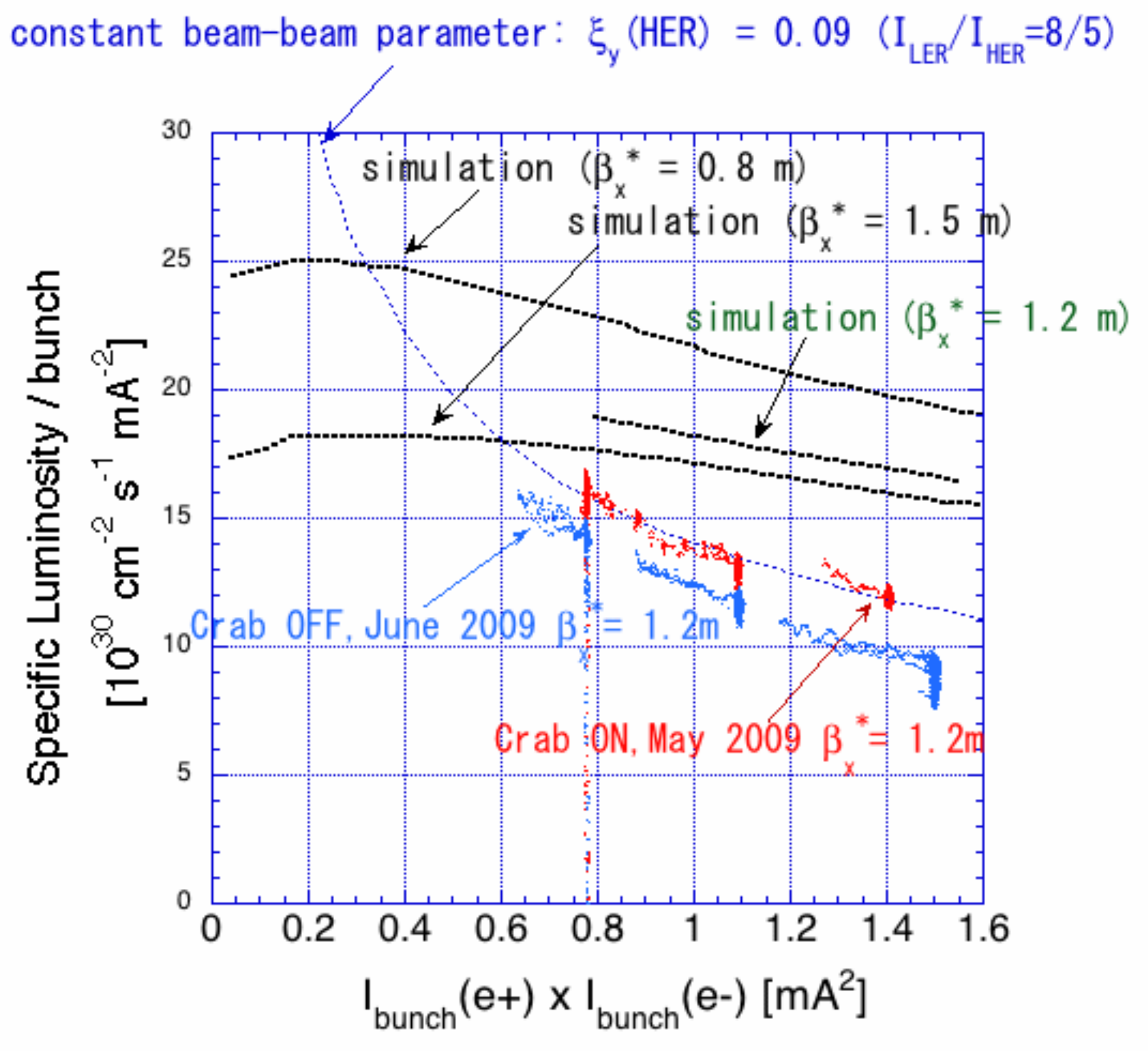}
\caption{Comparison of the specific luminosity per bunch with and without the crab cavities, as a function of the bunch current product of the two beams. The specific luminosity is defined as the luminosity divided by the bunch current product of the two beams, further divided by the number of bunches. In addition, three different lines from the beam--beam simulation are shown, corresponding to different values of the IP horizontal beta function, $\beta_x^*$. The simulations predicted that a smaller $\beta_x^*$ (smaller $\sigma_x^*$) would give a higher luminosity. Also shown is a line that corresponds to a constant vertical beam--beam parameter for an HER of 0.09, assuming that the bunch current ratio between LER and HER is 8\,:\,5. The data with crab cavities are aligned on this line, which means that the HER vertical beam--beam parameter, $\xi_y({\rm HER})$, is saturated at around 0.09.}
\label{SpecLum}
\end{figure}

%\subsection{Experience of beam operation}
%------------------------

\subsection{Tuning Method for Crab Cavity Parameters with Beams, and Beam Tuning with Crab Cavities}

\subsubsection{Crab voltage}
Prior to beam operation, calibration of the crab voltage was done by using the klystron output power and the loaded $Q$ values of the crab cavities without actual beams. The crab voltage was also calibrated by using beams. If a bunch passes by the crab cavity at the zero-cross timing of the crab RF voltage, the centre of the bunch receives no dipole kick. When the crab phase shifts from this condition, the bunch receives a net dipole kick from the cavity as in the case of a steering magnet. This dipole kick makes a closed orbit distortion (COD), and its size depends on the crab phase. From the CODs around the ring created by the crab cavity, the dipole kick angle can be estimated. By scanning the crab phase by more than $360^\circ$ and fitting the kick angle estimated at each data point as a function of the crab phase, the crab voltage can be determined. The crab voltage thus determined is consistent with that calibrated from the klystron power and the $Q$ value to within a few percent. From the crab phase scan and the fit, the phase shifter of the crab cavity system can also be calibrated. For the actual beam operation in the physics run mode, the crab voltages of both rings are scanned to maximize the luminosity, as shown below.

\subsubsection{Crab phase}
In principle, the crab phase should be set so that the centre of the bunch passes by at the zero-cross timing of the crab cavity. In this situation, the bunch receives no net dipole kick. This condition can be found by scanning the crab phase as described above; however, the method is rather time-consuming and so a easier approach is used in the usual operation. This alternative method involves searching by trial and error for the crab phase that brings no change in the COD between the crab on and off. Although there are two zero-cross phases, we can choose the correct phase by observing the phase of the COD. In the actual physics run, where high beam currents are needed, the crab phase is shifted by a certain  amount (typically $10^\circ$) to suppress the dipole oscillation observed at high-current crab collision. The COD induced by the net dipole kick from the crab cavity can be compensated for by employing steering magnets in the ring.

\subsubsection{Beam orbits at the crab cavities}
The beam loading for the crabbing mode increases linearly with a horizontal orbit displacement from the centre of the crab cavity. If the RF power to operate the cavity is too sensitive to the beam orbit, the cavity operation under the existence of the beams could be difficult.  To avoid this situation, we have chosen the loaded $Q$ value of the cavity to be ${Q}_{\rm L}  =$\,1--2\,$ \times 10^5$. With this relatively low $Q$ value, the RF power for the operation is relatively high (typically 100~kW at 1.4~MV); however, the RF power becomes less sensitive to the beam orbit (with a typical 20\% change per 1~mm orbit change). When we condition the cavity, we need a higher power; but with this $Q$ value, 200~kW is sufficient for conditioning the cavity up to 2~MV. In addition, we have developed an orbit feedback system to keep the horizontal beam orbit at the crab cavity stable \cite{CrabOF}. This system is composed of four horizontal steering magnets to make an offset bump for each ring, together with four beam position monitors (BPMs) for each ring to monitor the beam orbit at the crab cavity. The design system speed is 1~Hz, and the target accuracy of the orbit is within 0.1~mm. However, in the actual beam operation, we found that the beam orbit is stable enough even without the orbit feedback system. Therefore,  we usually do not use the orbit feedback system. At the beginning of the beam operation with the crab cavities, we searched for the field centre in the cavities by measuring the amplitude of the crabbing mode excited by beams when the cavities were detuned. In this search, the field centre of the HER crab cavity was found to be shifted by about 7~mm from the assumed centre position of the crab cavity.
A possible reason for this large displacement  is a misalignment of the cavity. We feel that there could be such a large misalignment, as precise alignment of the crab cavity to the cryostat is  very difficult.

\subsubsection{Luminosity tuning with crab cavities}
Luminosity tuning in general is  described above. Here we describe the method of  luminosity tuning related to the crab cavities. In the following, we discuss two tuning items: the crab V$_{c}$
(crab voltage) scan, and the tuning on the $x$--$y$ coupling at the crab cavities. For the crab V$_{c}$, the calibration can be done with a single beam as mentioned above; this, however, is not enough for the beam collision operation, since optics errors like those for the beta functions or the phase advance between the crab cavity and IP could shift the optimum crab V$_{c}$. In the actual tuning, we first tune the balance of the crab V$_{c}$ between the two rings. For this purpose, we employ a trick to change the crab phase slightly and observe the orbit offset at the IP. The IP orbit feedback system \cite{iBump} can detect the orbit offset at the IP precisely. Changing the crab phases of both rings by a certain amount (typically 10--15$^\circ$), we tune the balance of the crab V$_{c}$ between the two rings so that the IP orbit offset becomes the same for both rings. In this tuning, we rely on the accuracy of the phase shifter of the crab cavity system. Keeping this balance (the ratio of the crab V$_{c}$), we scan the crab V$_{c}$ for both rings and set the values that give the maximum luminosity. In our experience, the optimum set of the crab V$_{c}$ thus found is not much different from the calibrated values with the single beam. The difference is usually within 5\%.

The motivation for controlling the $x$--$y$ coupling at the crab cavities is to handle the vertical crabbing. In principle, the crab cavity kicks the beam horizontally; but if there is $x$--$y$ coupling at the crab cavity or if the crab cavity has some rotational misalignment, the beam could receive a vertical crab kick, and this may degrade the luminosity. The local $x$--$y$ coupling is expressed with four parameters, $R1$, $R2$, $R3$, and $R4$, as described above.  In the actual beam operation, these coupling parameters are scanned one by one to maximize the luminosity. We have  found that the tuning with these knobs has some effect on  the luminosity and that  the luminosity gain with the knobs is typically 5\%. We expected that $R2$ and $R4$ might have an effect on the luminosity, since these parameters are related to the vertical crab at the IP. In reality, however, there is no big difference in the effectiveness of the four parameters with respect to luminosity tuning.

%------------------------
%\subsection{Luminosity performance with crab cavities}
%------------------------

\subsection{Specific Luminosity With and Without the Crab Cavities}
Since the introduction of the crab cavities, we have made efforts~\cite{EPAC08,PTEPCom} to realize the beam--beam performance predicted by the beam--beam simulation. As a result of those efforts, we have achieved a relatively high beam--beam parameter of about 0.09, as shown in Table~\ref{tcrab2}. We have found  the correction of the chromaticity of the $x$--$y$ coupling at IP to be  effective in  increasing the luminosity~\cite{PTEPOv}. This correction increased the vertical beam--beam parameter from about $0.08$ to around $0.09$. However, even with this improvement, the  beam--beam parameter $0.09$  is still much lower than the  value of around $0.15$ predicted by simulation. We do not yet understand the cause of this discrepancy.

\begin{table}[hbt]%4
\tabcolsep 6pt
\begin{center}
\caption{Comparison of KEKB machine parameters with and without crab crossing.}
\vspace{2mm}
\begin{small}
\begin{tabular}{l cc cc c}
\hline\hline
& \multicolumn{2}{c}{\textbf{Jun  2010}}& \multicolumn{2}{c}{\textbf{Nov 2006}} & \\
& \multicolumn{2}{c}{With crab}& \multicolumn{2}{c}{Without crab} & \\
               &  LER       &   HER         &  LER      &     HER         &   Unit          \\ \hline
Energy         & 3.5        &   8.0         &  3.5      &     8.0         &   GeV       \\
Circum.  & \multicolumn{2}{c}{---\,3016\,---} & \multicolumn{2}{c}{---\,3016\,---}  &   m         \\
$I_{\rm beam}$ & 1637       &   1188        &  1662     &  1340           &   mA        \\
\#~bunches  & \multicolumn{2}{c}{---\,1585\,---} & \multicolumn{2}{c}{---\,1387\,---}  &             \\
$ I_{\rm bunch}$& 1.03       &   0.75        &  1.20     &  0.965          &   mA        \\
Avg. spacing   & \multicolumn{2}{c}{---\,1.8\,---} & \multicolumn{2}{c}{---\,2.1\,---}   &   m         \\
Emittance        &  18        &     24         &   18        &  24              &   nm        \\
$\beta_x^*$    &  120        &     120         &   59        &  56              &   cm        \\
$\beta_y^*$    &  5.9       &     5.9        &   6.5       &  5.9             &   mm        \\
Ver.~size\,@\,IP  &  0.94       &     0.94        &   1.8       &  1.8             & $\rm \mu m$     \\
RF voltage     &  8.0       &     13.0      &   8.0       &  15.0           &   MV        \\
$\nu_x$           & 0.506     &    0.511      &  0.505    & 0.509           &             \\
$\nu_y$          & 0.561      &    0.585     &  0.534    & 0.565           &
\\

$\xi_x$            & 0.127      &    0.102      &  0.117    & 0.071           &             \\
$\xi_y$            & 0.129      &    0.090      &  0.108    & 0.057           &             \\
Lifetime           &  133       &     200       &   110     &  180            & min.        \\
Luminosity     & \multicolumn{2}{c}{$\rm 2.108 \times10^{34}$}& \multicolumn{2}{c}{$ \rm 1.760  \times10^{34}$} &  $\rm /cm^2 /s$      \\
Lum/day        & \multicolumn{2}{c}{---\,1.479\,---}& \multicolumn{2}{c}{---\,1.232\,---} &  $\rm fb^{-1}$        \\
\hline\hline
\end{tabular}\end{small}
\label{tcrab2}
\end{center}
\vspace{-2mm}
\end{table}

Figure~\ref{SpecLum} compares the specific luminosity per bunch with the crab cavities on and off. The specific luminosity is defined as the luminosity divided by the bunch current product of the two beams, further divided by the number of bunches. If the beam sizes are constant with respect to the beam currents, the specific luminosity per bunch should be constant. As seen in Fig.~\ref{SpecLum}, the specific luminosity is not constant. This means that the beam sizes are enlarged as functions of the beam currents. In the experiment to obtain data in Fig.~\ref{SpecLum}, the number of bunches was reduced to 99 to avoid the possible effects of the electron clouds. In the usual physics operation, the number of bunches was 1585. For this experiment, the IP horizontal beta function, $\beta_x^*$, was changed from 0.8~m to 1.2~m to avoid the physical aperture problem and to increase the bunch currents. In the usual physics operation, the bunch current product was around 0.8~mA$^2$. The specific luminosity per bunch with the crab on is about 20\% higher than that with the crab off. Since the geometrical loss of the luminosity due to the crossing angle is calculated to be about 11\%, there is definitely some gain in the luminosity by the crab cavities other than recovery of the geometrical loss. However, the  effectiveness of the crab cavities is much smaller than in the beam--beam simulation, as can be seen in Fig.~\ref{SpecLum}. The beam--beam parameter is strictly constrained for some unknown reasons.

\subsection{Efforts to Increase Specific Luminosity with Crab Cavities}
Performance with the crab cavities has been considered very important not only for KEKB but also for SuperKEKB in the so-called high-current scheme. Therefore, we have been making every effort to understand the discrepancy between the beam--beam simulation and the experiments on beam--beam performance with the crab cavities. Although we have not identified the cause, we summarize our efforts as follows.

\begin{figure*}[thh]%5
\vspace{3mm}
\centering
\includegraphics*[width=140mm,clip]{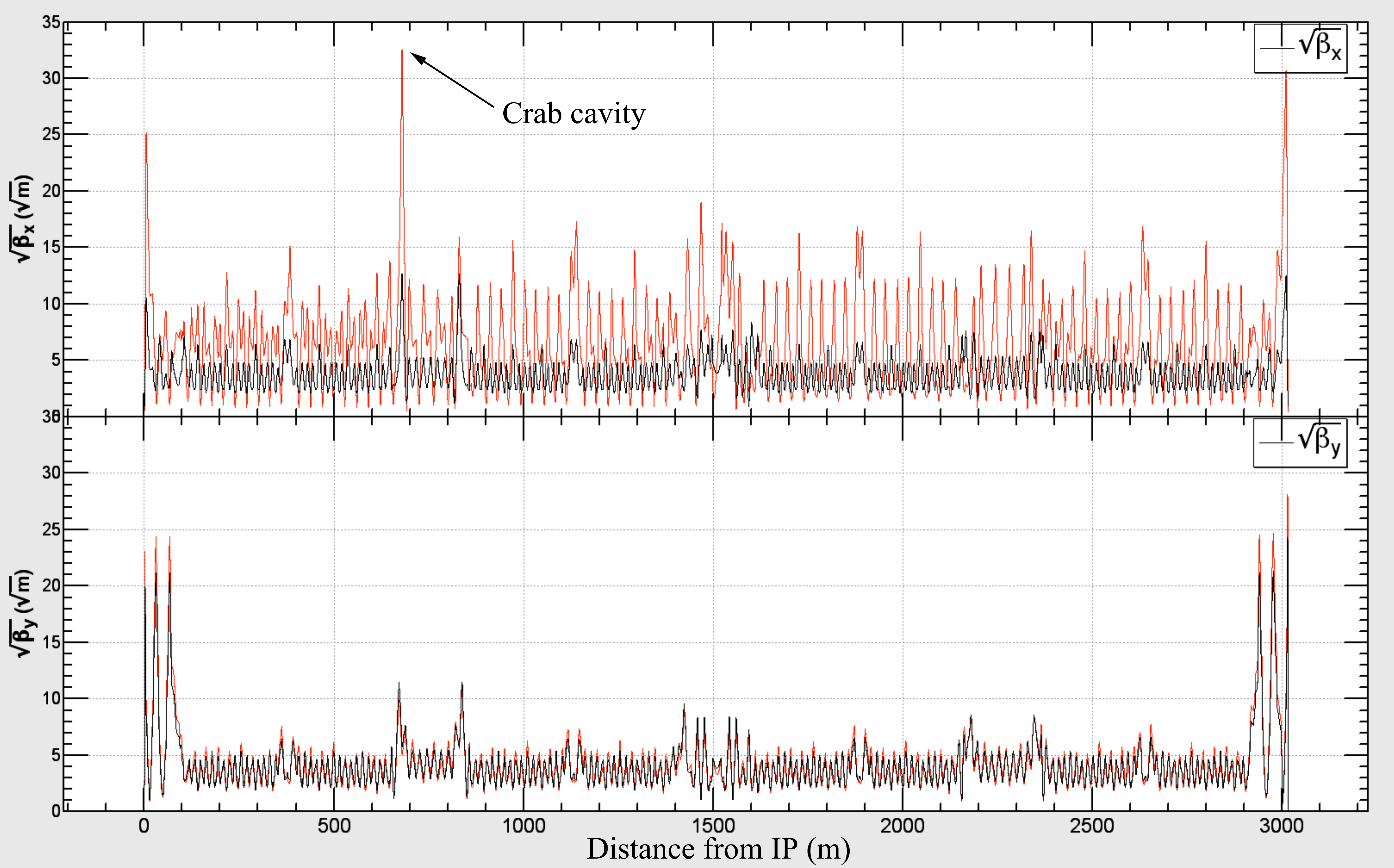}
\caption{Beating of  beta functions due to dynamic beam--beam effects in LER before we took  measures to counter this problem, with a  $\nu_x$ of 0.506 and an unperturbed beam--beam parameter $\xi_{x0}$ of 0.127. The red and black lines are the beta functions with  and without the dynamic beam--beam effects, respectively.}
\label{DynamicBeta}
\end{figure*}

\begin{figure}[tbh]%6
\vspace{3mm}
\centering
\includegraphics*[width=87mm,clip]{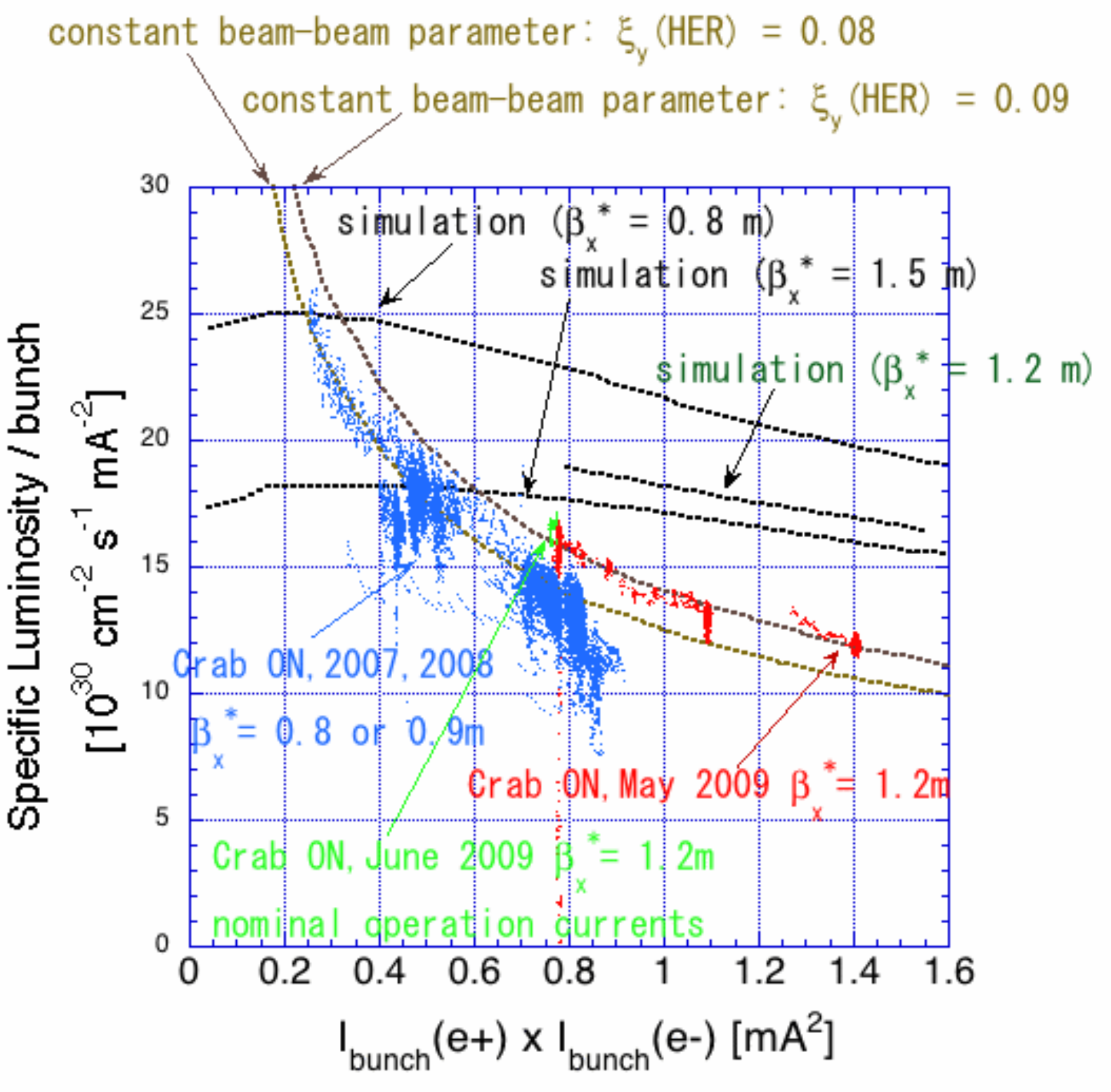}
\caption{Specific luminosity per bunch as a function of the bunch current product of the two beams,  for different values of $\beta_x^*$. In addition, three different lines from the beam--beam simulation are shown, corresponding to different values of the IP horizontal beta function,
$\beta_x^*$. The simulations predicted that a smaller $\beta_x^*$ (smaller $\sigma_x^*$) would give a higher luminosity. Also shown are lines that correspond to  constant vertical beam--beam parameters for HER values of 0.08 and 0.09, assuming that the bunch current ratio between LER and HER is 8\,:\,5. The data with crab cavities are aligned on those lines, which means that the HER vertical beam--beam parameter, $\xi_y({\rm HER})$, was saturated at around 0.08 or 0.09. In the experiment, we found that the luminosity did not depend on the IP horizontal beta functions $\beta_x^*$, in contrast to the simulation. The data with $\beta_x^* = 0.8$ or 0.9~m (blue dots) was collected before we introduced the skew-sextupole magnets. The data obtained after  introduction of the skew-sextupoles (green and red dots) are aligned on the line corresponding to $\xi_y({\rm HER})= 0.09$. This means that the maximum beam--beam parameter increased from 0.08 to 0.09 because of the skew-sextupoles. We changed $\beta_x^*$ from 0.8 or 0.9~m to 1.2~m to increase the bunch currents by mitigating the physical aperture problem at the crab cavities and to be able to compare the data with simulations at a higher bunch current region. Even upon  solving the physical aperture problem,  a large discrepancy persisted between the simulation and the experiment.}
\label{SpecLum2}
\end{figure}

\subsubsection{Short beam lifetime related to physical aperture around crab cavities}

In  beam operation with the crab cavities, we encountered a situation where we could not increase the bunch current of one beam due to poor beam lifetime of the other beam. We took this issue seriously and made efforts to overcome it, since it  is a possible cause of degradation of the beam--beam performance with the crab crossing. We were able to identify the process responsible for the lifetime decrease: dynamic beam--beam effects, i.e.\ the dynamic beta effect and the dynamic emittance effect. Since the horizontal tune of KEKB is very close to a half-integer, the effects are very large.  In Fig.~\ref{DynamicBeta}, the beta functions around the LER ring are depicted with and without the dynamic beam--beam effect before we solved the problem. The horizontal beta function around the crab cavity becomes very large.  Here, the horizontal tune was 0.506 and the unperturbed horizontal beam--beam parameter was around 0.127, with the operation bunch current of HER. Without the beam--beam perturbation, the horizontal beta functions at the IP and at a quadrupole magnet next to the crab cavity were 0.9~m and 161~m, respectively. With the beam--beam effect, the beta functions were calculated to be 0.138~m and 1060~m at the IP and the quadrupole magnet, respectively. To meet the crab condition, the horizontal phase advance between the crab cavity and the IP was chosen to be $\pi/2$ times an odd integer. With this phase advance, the horizontal beta function becomes very large around the crab cavity.  Also, due to the dynamic beam--beam effect, the horizontal emittance ($\varepsilon_x$) was enlarged from 18~nm to around 52~nm. In this situation, we have found that the horizontal beam size around the crab cavity is very large (typically 7~mm) at the operation bunch currents, and the physical aperture there is only around $5\sigma_x$. Therefore, the physical aperture around the crab cavities could seriously affect the beam lifetime.  The same problem is also observed at HER, although the effect is less serious because the horizontal tune of HER is further away from the half-integer than in the LER case.

To mitigate this problem, we have taken several counter-measures. In the original optics of LER,  the horizontal beta function around the crab cavity took the local maximum value not at the crab cavity but at the quadrupole magnets closest to the crab cavity. To satisfy the crab condition, the horizontal beta function at the crab cavity should be set at the target value. If we can decrease the beta function at the quadrupole magnet while keeping the beta function at the crab cavity unchanged, we can widen the physical acceptance around the crab cavity. During the summer shutdown of  2008, we changed the optics around the crab cavity by adding some power supplies for the quadrupole magnets and changing the wiring of the power supplies.
As a result, the horizontal beta function at the quadrupole magnets next to the crab cavity was reduced to the same value as at the crab cavity. Before this change, the horizontal beta function at the quadrupoles was about twice as large as that at the crab cavity.
With this change, the beam lifetime problem was alleviated to some extent; however, when we increased the bunch currents beyond the usual operation values, the lifetime problem appeared again. To investigate the specific luminosity at higher bunch currents, we decided to increase the horizontal beta function at the IP. By enlarging the IP beta function, we can lower the beta function at the crab cavity and enlarge the physical acceptance. We enlarged $\beta_x^*$ from 0.8~m or 0.9~m to 1.2~m or 1.5~m.  With this change, we were able to increase the bunch currents up to the value shown in Fig.~\ref{SpecLum}, and the discrepancy between the simulation and the experiment became more evident. Figure~\ref{SpecLum2} shows a comparison of the specific luminosity with different values of $\beta_x^*$. In the beam--beam simulations,  as shown in the figure, the specific luminosity with $\beta_x^*= 0.8$~m is much higher than that with $\beta_x^*= 1.5$~m. In the experiment, however, such a change in $\beta_x^*$ did not make any difference to the specific luminosity. The specific luminosity with $\beta_x^* = 0.8$~m or 0.9~m in Fig.~\ref{SpecLum2} is lower than that with $\beta_x^* = 1.2$~m. This is because the data with $\beta_x^* = 0.8$~m or 0.9~m was taken before the introduction of the skew-sextupole magnets.  In Fig.~\ref{SpecLum2}, the specific luminosity with the nominal operation bunch currents is also shown (as green dots) for reference.
In addition to these counter-measures for the lifetime problem, we also tried to raise the crab voltage. If this were successful, we could have lowered the horizontal beta function at the crab cavity while keeping $\beta_x^*$ the same. We tried to operate the He refrigerator with lower pressure to lower the He temperature. From the data in the R\&D stage, it was expected that we can operate the crab cavity stably with a higher voltage, if the He temperature was lowered. We actually succeeded in lowering the He temperature from 4.4~K to 3.85~K in April 2009. Nevertheless,  the maximum crab voltage turned out to be unchanged even with this lower He temperature. Therefore, we gave up this trial.

With these counter-measures in place, we also expected to improve the specific luminosity by solving the lifetime problem, since we sometimes encountered a situation where we could not move some machine parameter, such as a horizontal orbital offset at IP, in the direction giving a higher luminosity because of poor beam lifetime. We found, however, that the lifetime problem has almost nothing to do with the specific luminosity, except in the  region of high bunch current where the lifetime problem was particularly serious.

For the short lifetime problem, we have developed another counter-measure of e$^+$/e$^-$ simultaneous injection.
The injector linac is shared by four accelerators: two are the KEKB rings, and the other two are the PF ring and an SR ring called PF-AR.  Before the successful introduction of the simultaneous injection scheme, there were four injection modes corresponding to the four rings. Switching from one mode to another took from about  30~s to around  3 minutes. The idea of  simultaneous injection is to switch the injection modes pulse-to-pulse. In the period of KEKB operation, we successfully implemented simultaneous injection for three rings (the two KEKB rings and the PF ring) \cite{PTEPInj,PTEPBT}.
With this new injection scheme,  beam operation with shorter beam lifetime became possible. However, as mentioned above, we found that the lifetime problem has almost nothing to do with the specific luminosity, even though the machine parameter scan at KEKB has become much faster with constant beam currents stored in the rings and it has become possible to find better machine parameters much faster than before.

\subsubsection{Synchro-betatron resonance}
In the KEKB operation, we found that the synchro-betatron resonance of ($ 2\nu_x + \nu_s = \textrm{integer}$) or ($2\nu_x + 2\nu_s = \textrm{integer}$) seriously affects KEKB performance. The nature of the resonance lines was examined in detail during the
machine study on crab crossing.
We found that the resonances affect (1) single-beam lifetime, (2) single-beam beam sizes (in both  horizontal and vertical directions), (3) two-beam lifetime, and (4) two-beam beam sizes (in both horizontal and vertical directions); moreover, the effects are beam-current dependent. The effects lower the luminosity directly or indirectly through  beam size blow-up,  beam current limitation due to poor beam lifetime, or a smaller variable range of the tunes.
The strength of the resonance lines can be weakened by suitably choosing  a set of sextupole magnets. KEKB adopted the non-interleaved sextupole scheme to minimize nonlinearity of the sextupoles. LER and HER have 54 pairs and 52 pairs of sextupoles, respectively. With so many degrees of freedom in the number of the sextupoles, optimization of the sextupole setting is not an easy task even with current computing power. Prior to the beam operation, the candidates for the sextupole setting are searched for by  computer simulation. Usually, dynamic aperture and an anomalous emittance growth are optimized on the synchro-betatron resonance.
%
%At KEKB an efficient method of optimization has been developed by using Temperature Parallel %Simulated Annealing (TPSA) method~\cite{AM}.
A setting of sextupoles that  gives good performance in the computer simulation does not necessarily bring good performance in the real machine, and most  candidates for the sextupole setting do not yield satisfactory performance. When we changed linear optics, we usually  needed to try many candidates  before finally obtaining a setting with adequate performance. The single-beam beam size and  beam lifetime are criteria for sextupole performance. Alternatively, as an easier method for estimating sextupole performance, a beam loss was observed when the horizontal tune was jumped down across the resonance line.  The resonance line in HER is stronger than that in LER, since there is no local chromaticity correction in HER. In usual operation, we could operate the machine with the horizontal tune below the resonance line in the LER case, whereas we could not lower the horizontal tune of HER below the resonance line. The beam--beam simulation predicts a higher luminosity with the lower horizontal tune in HER. To weaken the strength of the resonance line in HER, we tried to change the sign of $\alpha$ (momentum compaction factor). Since $\nu_s$ is negative for positive $\alpha$, the resonance is a sum resonance ($ 2\nu_x + \nu_s = \textrm{integer}$). By switching the sign of $\alpha$, we can change it to a difference resonance ($2\nu_x - \nu_s = \textrm{integer}$). The trial was undertaken in June 2007; it was successful and we were able to lower the horizontal tune below the resonance. However, when we tried the negative $\alpha$ in LER, an unexpectedly large synchrotron oscillation due to the microwave instability occurred. Because of this oscillation, we gave up the trial of the negative-$\alpha$ optics. So far, we have no definite conclusion about the effect of the synchro-betatron resonance on the specific luminosity.

\subsubsection{Machine errors}

The method of luminosity tuning was described earlier. In the conventional method of tuning at KEKB, most parameters (except for those optimized by observing their own observables) are scanned one at a time just by observing the luminosity and  beam sizes. One possible explanation for the low specific luminosity is that we have not yet reached an optimum parameter set, due to the  parameter space being too wide. As a more efficient method of parameter search, we introduced in autumn 2007 the downhill simplex method for 12  parameters, consisting of the $x$--$y$ coupling parameters at the IP as well as the vertical dispersions at IP and their slopes, which from the experience of KEKB operation are very important for luminosity tuning. These 12 parameters can be searched for at the same time with this method. We have been using the method ever since. Nevertheless, even with this method we have not managed to achieve an improvement in specific luminosity, although the speed of parameter search seems to be rather enhanced.

Another possible reason for not being able to achieve a higher luminosity with the above tuning method is the side effects of the large tuning knobs. Although machine errors can be compensated for by using the tuning knobs, too-large tuning knobs bring side effects that  would degrade the luminosity. Therefore, if the machine errors are too large, the luminosity predicted by the simulation cannot be achieved by using the usual tuning knobs. We have actually confirmed that large tuning knobs on the $x$--$y$ coupling at the IP can degrade  single-beam performance. The question is how large are the  machine errors that  exist at KEKB. According to the simulation, with reasonable machine errors such as misalignments of magnets and BPMs,  offsets of BPMs, and  strength errors of the magnets,  large errors of the $x$--$y$ coupling or the dispersion at IP are not created, as the luminosity cannot be recovered by the knobs because of their side effects.  One possibility would be the error related to the detector solenoid. The Belle detector is equipped with the 1.4~T solenoid. The field is locally compensated for by the compensation solenoid magnets installed near the IP, so that the integral of the solenoid field is zero on both sides of the IP. The remaining effects of the solenoid field are compensated for by the skew-quadrupole magnets located close to the IP.  If the compensation is not enough (or if it over-compensates),  a large error of the $x$--$y$ coupling would remain. Although there is no direct evidence that the compensation of the Belle solenoid is not enough, the effect of the Belle solenoid on the luminosity has been doubted, as for the beam-energy dependence of the luminosity.  KEKB was designed to operate on the $\Upsilon (4{\rm S}) $ resonance ($E_{\rm CM} =  10.58$~GeV). KEKB was also operated on $\Upsilon (1{\rm S}) $ ($E_{\rm CM} =  9.46$~GeV), $\Upsilon (2{\rm S}) $ ($E_{\rm CM} =  10.02$~GeV), and $\Upsilon (5{\rm S})$ ($E_{\rm CM} =  10.87$~GeV). We found that the luminosity on $\Upsilon (5{\rm S}) $ is almost the same as that on $\Upsilon (4{\rm S}) $. However, the luminosity on $\Upsilon (1{\rm S}) $ and $\Upsilon (2{\rm S}) $ is lower than that on $\Upsilon (4{\rm S}) $ by  about 50\% and 20\%, respectively. The design beam energy of KEKB is that of $\Upsilon (4{\rm S}) $,  and the $x$--$y$ coupling due to the Belle solenoid is compensated for completely at this design energy. When we change the beam energy, we do not change the strength of the Belle solenoid and the compensation solenoids. Thus, the $x$--$y$ coupling correction for the Belle solenoid is not complete on the resonance other than $\Upsilon (4{\rm S}) $, and the luminosity would be affected by the remaining $x$--$y$ coupling. To investigate this issue, a machine study was done on $\Upsilon (2{\rm S}) $  in October 2009 with the Belle solenoid and the compensation solenoid  tracked to the beam energy. Contrary to the initial expectation, the luminosity in this condition was even worse than the usual 2S run. We gave up this trial after about two days, since the Belle experiment could not use the data with the different strength of the detector solenoid. Therefore, the correlation between the detector solenoid and the luminosity was not confirmed in this experiment.

We also tried to measure the $x$--$y$ coupling at the IP directly by using the injection kicker magnets and the BPMs around IP. Although some data showed a very large value of the $x$--$y$ coupling at IP, we have obtained no conclusive results because of the  poor accuracy of the measurements.

\subsubsection{Vertical emittance in a single-beam mode}
 The beam--beam simulation showed that the attainable luminosity depends strongly on the single-beam vertical emittance. If the actual vertical emittance is much larger than the assumed value, it could create the discrepancy. We carefully checked the calibration of the beam size measurement system. We found some errors in the calibration of the HER beam size measurement system, and the actual vertical emittance was somewhat smaller than the value  considered before. However, the latest values of the global $x$--$y$ coupling of the two beams are around 1.3\%, and these values of the coupling do not explain the discrepancy in specific luminosity between the experiment and the simulation shown in Fig.~\ref{SpecLum2}, where the $x$--$y$ coupling in the simulation is assumed to be~1\%.

\subsubsection{Vertical crabbing motion}
The vertical crab at the IP could degrade the luminosity. It may be created by some errors related to the crab kick, such as a misalignment of the crab cavity and the local $x$--$y$ coupling at the crab cavity. The $x$--$y$ coupling parameters at the crab cavities give a tuning knob to adjust the vertical crab at the IP. By such tuning, we can eliminate the vertical crab at the IP even if it is created by other sources such as a  misalignment of accelerating cavities. Nevertheless, the tuning of these parameters does not suffice to increase the luminosity, as discussed above.

\subsubsection{Off-momentum optics}
It has been shown by beam--beam simulation that the chromaticity of the $x$--$y$  coupling at the IP could reduce the luminosity largely through the beam--beam interaction, if the residual chromatic coupling is large \cite{KO1,DZ}. While even an ideal lattice has such a chromatic coupling, the alignment errors of the sextupole magnets could create a large chromatic coupling. It has been thought that this kind of chromatic coupling might be one factor responsible for  the serious luminosity degradation with crab crossing. Parallel to trials for measuring such chromatic couplings directly, we introduced tuning knobs to control them. For this purpose, we installed 14 pairs of skew-sextupole magnets (10 pairs for HER and 4 pairs for LER) in early 2009. The maximum strength of the magnets (bipolar) is $K_2 \sim 0.1/\textrm{m}^2$ for HER and $K_2 \sim 0.22/\textrm{m}^2$ for LER.  By using these magnets, the tuning knobs were introduced to the beam operation at the beginning of May 2009. The luminosity gain due to these knobs is about 15\%. Even with the improvement in the luminosity obtained by the use of skew-sextupole magnets, there is still a large discrepancy between the experiment and the simulation.

\subsubsection{Fast noise}
Fast  noise could lead to a loss in the luminosity. According to the beam--beam simulation, the allowable phase error of the crab cavities for $N$-turn correlation is $0.1 \times \sqrt{N}$ degrees. On the other hand, the measured error in the presence of the beams was less than $\pm 0.01$ degree for fast fluctuation  (1~kHz or faster) and less than $\pm 0.1$ degree for slow fluctuation (from 10 to several hundred hertz). The measured phase error is then much smaller than the allowable values given by the beam--beam simulation. Besides the noise from the crab cavities, any fast noise could degrade the luminosity.  For example, a phenomenon we encountered  in 2005 is that the luminosity depends on the gain of the bunch-by-bunch feedback system. With a higher gain of about 6~dB, the luminosity decreased by about 20\% \cite{PTEPMon}. This seems to indicate that some noise in the feedback system degraded the luminosity; this phenomenon disappeared, however, after the system adjustment, which included replacement of an amplifier for the feedback system. Although we confirmed that some artificially strong noise introduced to the crab cavities or to the feedback system can decrease the luminosity \cite{CrabNoise}, there is no evidence that the achievable luminosity at KEKB was limited by  fast noise.

\begin{figure}[htb]%7
\centering
\vspace{-2mm}
\includegraphics*[width=75mm]{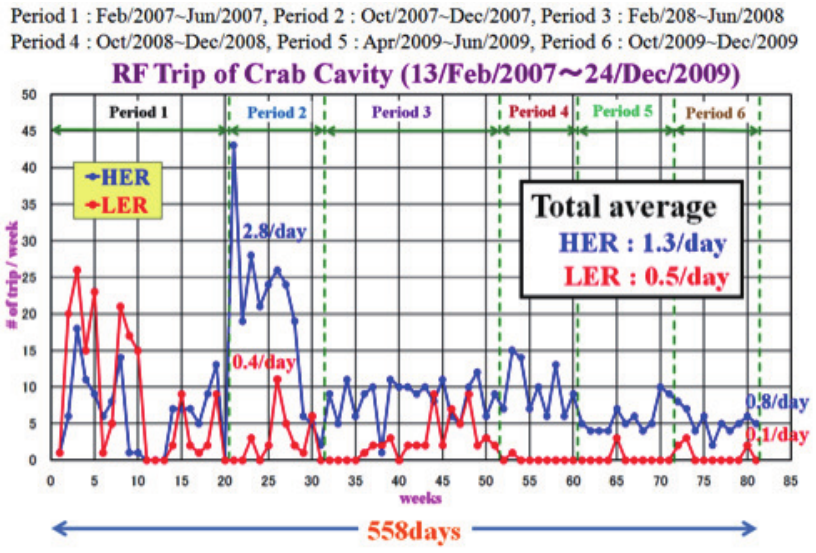}
\caption{Trip rate of crab cavity system.}
\label{f7}
\vspace{-3mm}
\end{figure}

\section{EXPERIENCE OF CRAB CAVITY OPERATION WITH BEAMS}
The initial goal of the beam study of the crab cavities was to show that the high beam--beam parameters predicted by the simulation are actually achievable in a real machine. This study could be done with relatively low beam currents  and with a smaller number of bunches. High beam current operation of the crab cavities was the second priority, as the tolerance of the crab cavities for high beam currents was unknown. However, they have been working much more stably than  initially expected  and are currently being used in the usual physics run. Figure~\ref{f7} shows a history of the trip rate of the crab cavities. Period~1 was a dedicated machine time for the study of the crab cavities and the crab crossing. In most  cases the beam currents are rather low, typically 100~mA for LER  and 50~mA for HER. Around the sixth week, the maximum attainable kick voltage of the LER crab cavity dropped suddenly from 1.5~MV to about  1.1~MV for an unknown reason. In the middle of this period, we had to warm up the system to  room temperature to recover from frequent trips of LER crab cavities. It was also expected that the performance degradation of the LER crab cavity would be recovered with the warm-up; however, the performance was not improved and this problem remains unsolved since then.
In the summer shutdown following Period~1, the cavities were warmed up again to room temperature. From Period~2, the use of the crab cavities in the usual physics run started. At the beginning of this period, we were troubled with frequent trips of the HER crab cavity. This problem was solved by lowering the crab voltage, which was possible by enlarging the horizontal beta function at the crab cavity and by RF conditioning. In the winter shutdown following Period~2, the cavities were warmed up once again to room temperature. During Period~3, the trip rate of the HER crab cavity seemed to be more or less stable, while that of the LER crab had a tendency to increase slowly after the warm-up. Generally speaking, the HER crab cavity shows a higher trip rate than that of LER, corresponding to the higher crab voltage as shown in Table~\ref{l2ea4-t1}. It seems that the situation with the trip rate has reached a more or less steady state and will continue in a similar manner from now on. As for the causes of the trips, most of the HER cases are attributable to breakdowns of superconductivity due to discharge in the cavity; causes for the LER cavity include discharge in the coaxial coupler or at the input coupler.

\end{document}